\documentclass{elsart1p}

\usepackage{epsfig}

\usepackage{amssymb}

\def\cc{$^{12}$C+$^{12}$C\ }
\def\AA{nucleus-nucleus\ }

\def\bbox#1{\mbox{\boldmath $#1$}}
\begin{document}
\begin{frontmatter}

% Title, authors and addresses
% use the thanksref command within \title, \author or \address for footnotes;
% use the corauthref command within \author for corresponding author footnotes;
% use the ead command for the email address,
% and the form \ead[url] for the home page:
\title{Mean-field description of the nucleus-nucleus optical potential\thanksref{label1}}
\thanks[label1]{Research supported, in part, by the Natural Science Council of Vietnam.}

\author{Dao T. Khoa and Do Cong Cuong}
\address{Institute for Nuclear Science and Technique, VAEC,
P.O. Box 5T-160, Nghia Do, Hanoi, Vietnam.}

\begin{abstract} A new finite-range representation of the
JLM effective nucleon-nucleon interaction is suggested based on the CDM3Y
density dependent functional and M3Y-Paris interaction. The density dependence
has been carefully adjusted at each energy so that the nucleon optical
potential (OP) given by the new density dependent interaction in the
Hartree-Fock calculation of nuclear matter closely matches the JLM nucleon OP
given by the Brueckner-Hatree-Fock calculation. The new interaction has been
used in the double-folding model to calculate the OP for the elastic
$^{6}$Li,$^{6}$He,$^{12}$C + $^{12}$C scattering at different energies.
\end{abstract}

%\begin{keyword}
% keywords here, in the form: keyword \sep keyword
% PACS codes here, in the form: \PACS code \sep code
%\PACS
%\end{keyword}
\end{frontmatter}

\begin{figure}[h]
\centering\vspace*{1.3cm}%\hspace*{-2cm}
 \mbox{\epsfig{file=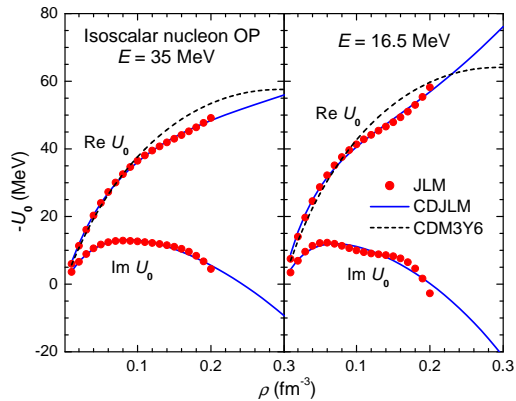,height=6cm}}\vspace*{-2cm}
\caption{$U_0(E,\rho)$ at $E=16.5$ and 35 MeV, given by the HF calculation
(\ref{e2}) using the CDJLM interaction, in comparison with the JLM results
\cite{Je77} and Re~$U_0(E,\rho)$ given by the CDM3Y6 interaction \cite{Kh97}.}
\label{f1}
\end{figure}

\begin{figure}[htb]\vspace*{-2cm}
\begin{minipage}[t]{80mm}
\hspace*{0.5cm}\vspace*{0cm}
\mbox{\epsfig{file=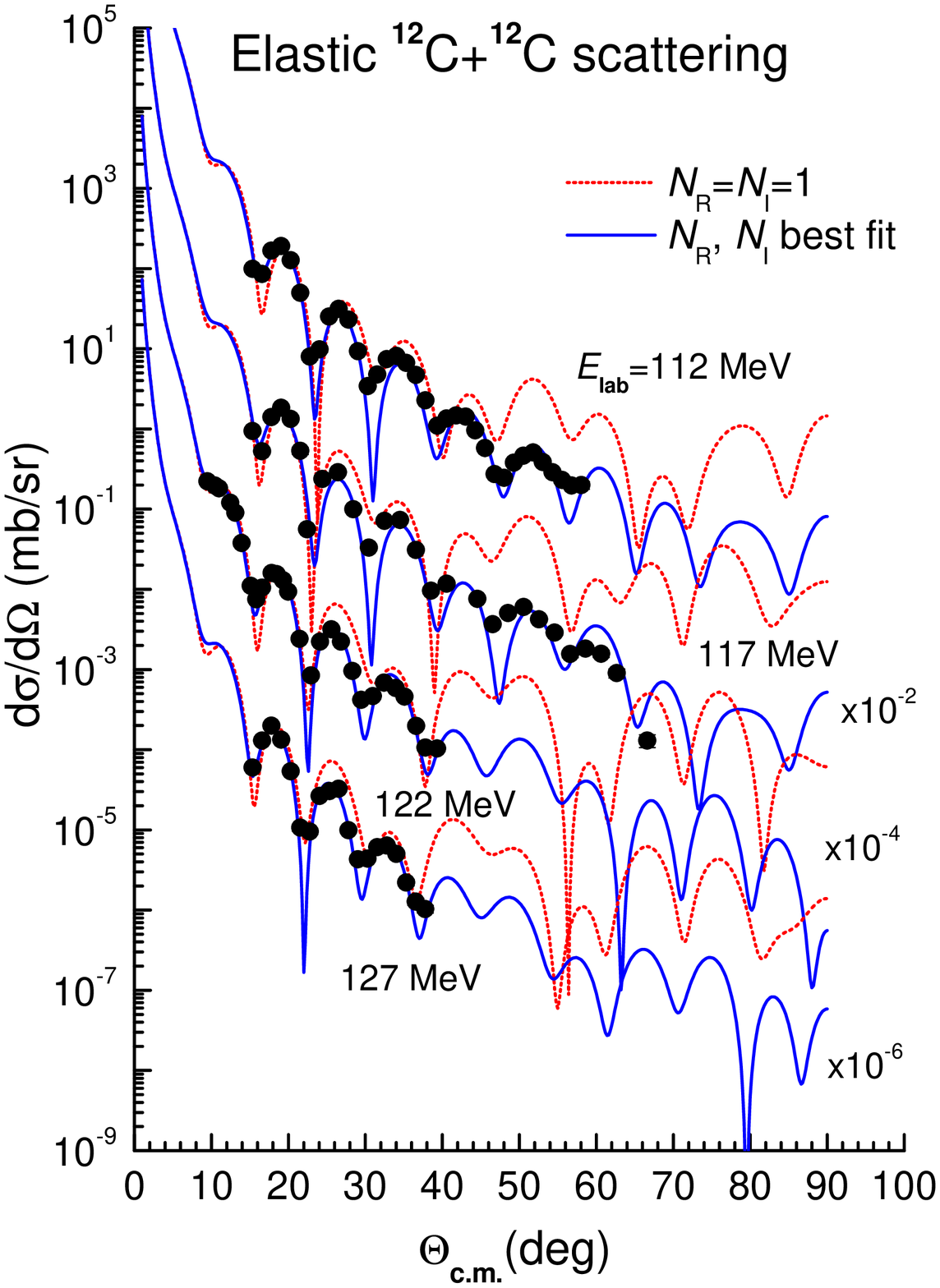,height=8.5cm}}
\end{minipage}
\hspace{\fill}
\begin{minipage}[t]{75mm}
\hspace*{-2cm}\vspace*{0cm}
\mbox{\epsfig{file=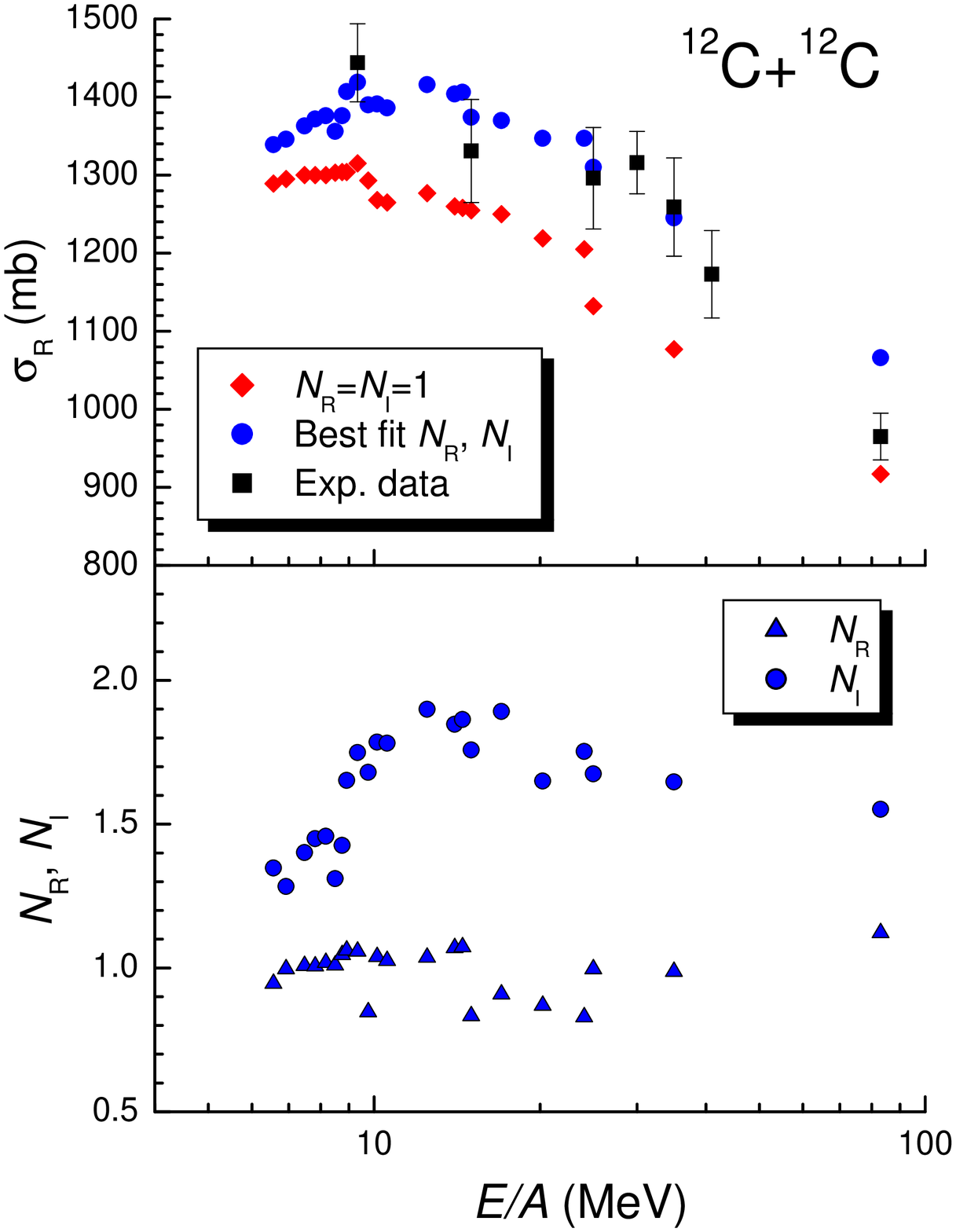,height=8.5cm}}
\end{minipage}
\vspace*{-1cm} \caption{\small Description of the elastic scattering and total
reaction cross section for \cc system given by $U_{\rm F}=N_{\rm R}V_{\rm
F}+iN_{\rm I}W_{\rm F}$. The best-fit $N_{\rm R}$ and $N_{\rm I}$ coefficients
are shown in the lower right panel.} \label{f2}
\end{figure}

\begin{figure}[htb]\vspace*{-0.5cm}
\begin{minipage}[t]{80mm}
\hspace*{0.5cm}\vspace*{0cm} \mbox{\epsfig{file=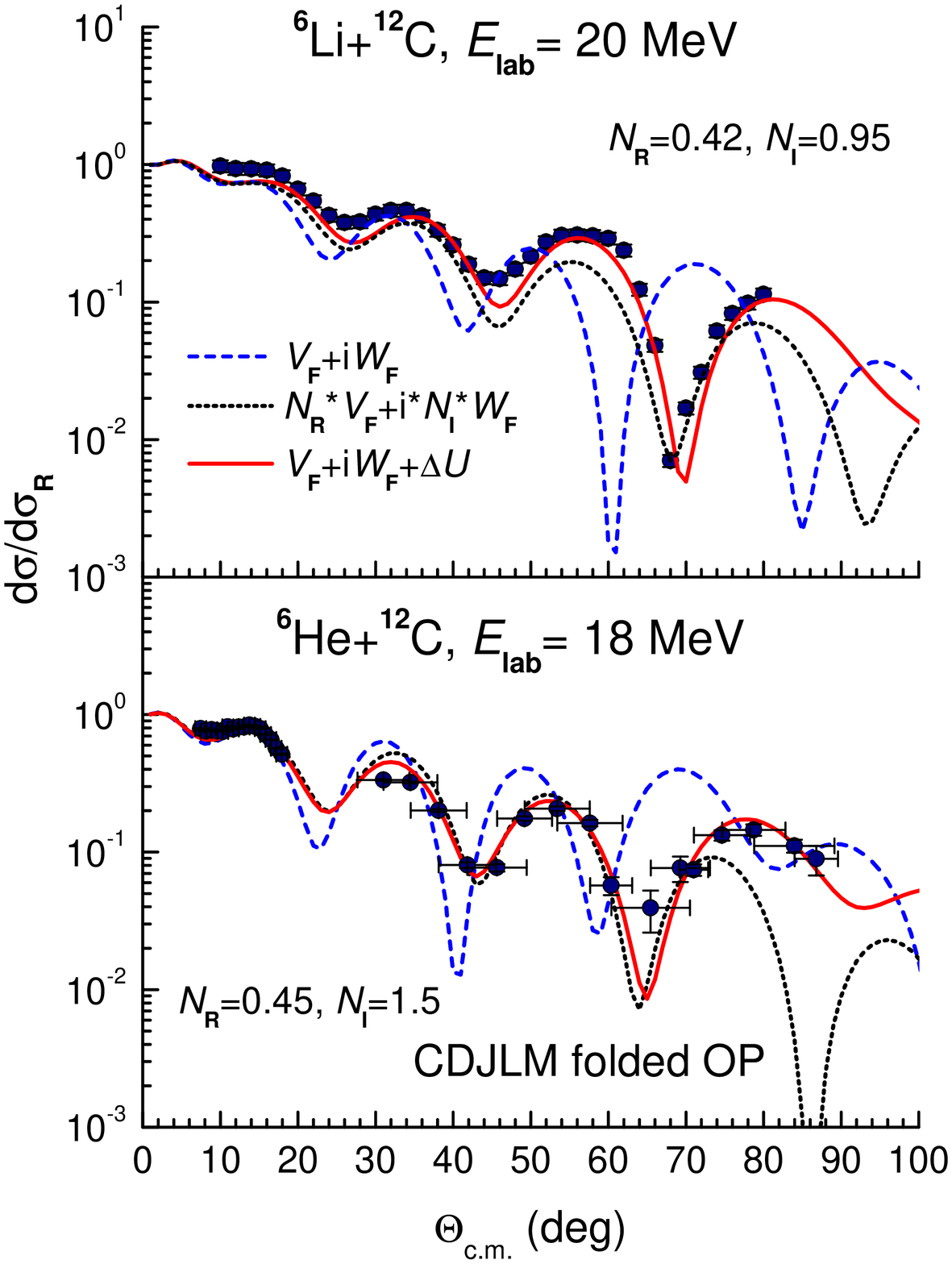,height=8cm}}
\end{minipage}
\hspace{\fill}
\begin{minipage}[t]{75mm}
\hspace*{-2cm}\vspace*{0cm} \mbox{\epsfig{file=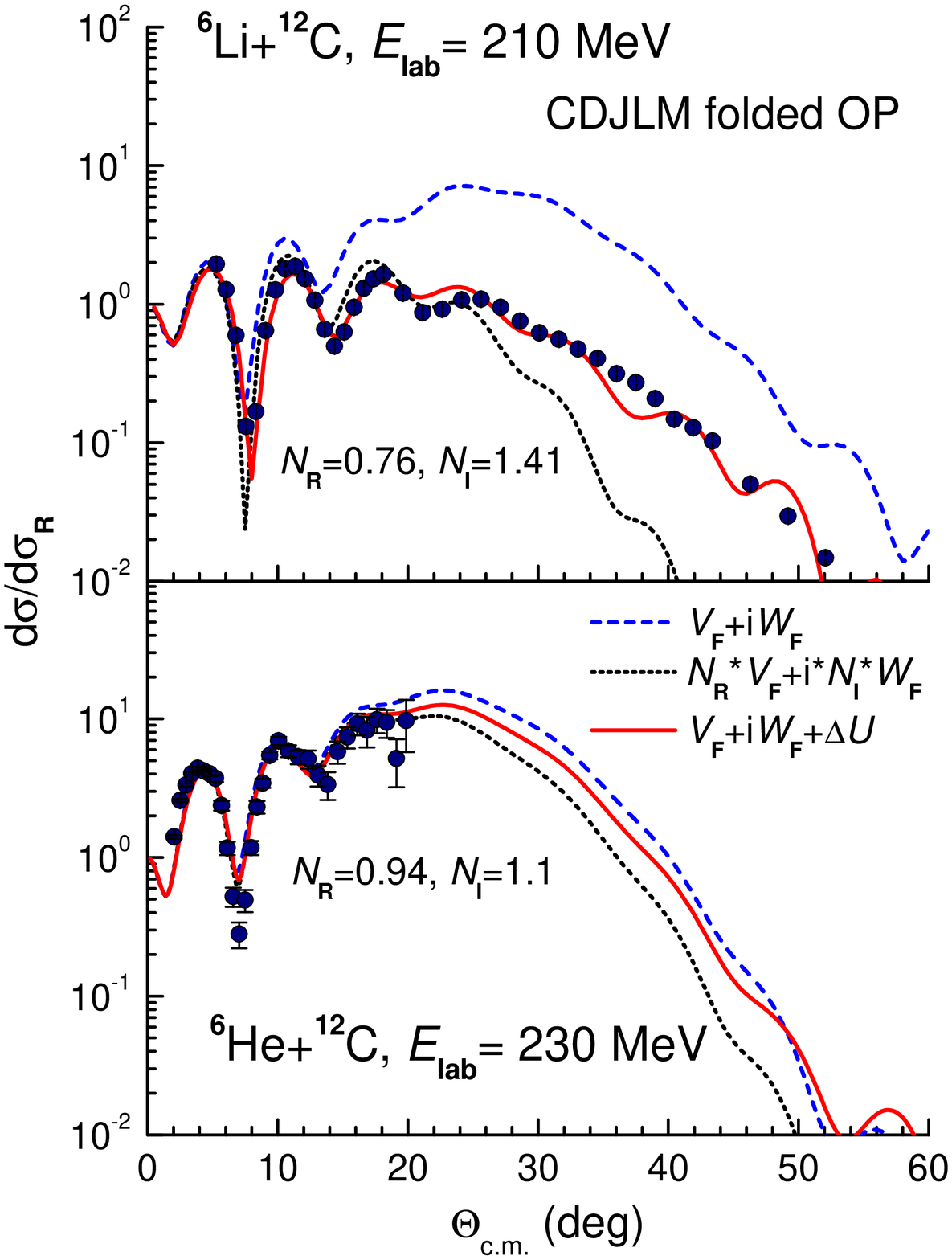,height=8cm}}
\end{minipage}
\vspace*{-1cm} \caption{\small Description of the elastic $^{6}$Li,$^{6}$He +
$^{12}$C scattering at 3 and 35 MeV/nucleon given by the folded $U_{\rm F}$
potential either renormalized or added by a complex (surface-peaked)
polarization potential $\Delta V$.} \label{f3}
\end{figure}

% main text
%\section{Deadline for submission}
%\label{}
Although the optical potential between two composite nuclei is a complicated
many-body problem due to the heavy-ion (HI) collision dynamics, an approximate
microscopic approach can be formulated based on the reaction theory by Feshbach
\cite{Fe92}
\begin{equation}
 U=V_{00}+\lim_{\epsilon \rightarrow 0}\sum_{\alpha\alpha'}
 {\kern -1pt ^\prime}
 V_{0\alpha}\Big({1\over E-H+i\epsilon}\Big)_{\alpha\alpha'} V_{\alpha'0}=
 V_{00} + \Delta V.
\label{e1}
\end{equation}
Here $V_{00}$ is the first-order term of the \AA OP and $\alpha$ stands for a
given projectile-target partition. The primed sum runs over all the excited
partitions excluding elastic channel ($\alpha=0$). The higher-order dynamic
polarization potential $\Delta V$ arises from couplings to all open nonelastic
channels ($\alpha\neq 0$) and is commonly believed to be the main source of the
\emph{absorption} in the HI system. However, when $V_{00}$ is evaluated within
the double-folding model (DFM) using a G-matrix interaction, it will be
\emph{complex} ($V_{00}\to U_{\rm F}=V_{\rm F}+iW_{\rm F}$) if the original
G-matrix is complex. $W_{\rm F}$ gives absorption of the \emph{mean-field}
origin which is due to a \emph{finite} mean-free path of nucleon in the medium
and \emph{not} caused by the nonelastic reaction channels. The total absorption
must then be given by $W_{\rm F}$+Im~$\Delta V$. To have an accurate mean-field
prediction of $U_{\rm F}$, we have constructed a new \emph{complex} density
dependent NN interaction using the Brueckner Hartree-Fock results for the JLM
nucleon OP \cite{Je77}. Namely, the \emph{isoscalar} complex nucleon OP in
nuclear matter is determined from the Hartree-Fock (HF) matrix elements of the
effective NN interaction between the incident nucleon and those bound in Fermi
sea as
\begin{equation}
  U_0(E,\rho)=\sum_{j\le k_F}[<\bbox{kj}|u_{\rm D}(E,\rho)|\bbox{kj}>+
  <\bbox{kj}|u_{\rm EX}(E,\rho)|\bbox{jk}>]. \label{e2}
\end{equation}
Here $k_F=[1.5\pi^2\rho]^{1/3}$ and $k$ is the momentum of the incident nucleon
which must be determined self-consistently as
 $k=\sqrt{2m[E-{\rm Re}~U_0(E,\rho)]/\hbar^2}$. We have used in Eq.~(\ref{e2})
two different CDM3Y functionals \cite{Kh97} to construct the \emph{real} and
\emph{imaginary} parts of the isoscalar density dependence of the interaction
\begin{equation}
 F_x(E,\rho)=C_x(E)[1+\alpha_x(E)\exp(-\beta_x(E)\rho)-\gamma_x(E)\rho],
 \label{e3}
\end{equation}
so that the real ($x=V$) and imaginary ($x=W$) parts of $u_{\rm D(EX)}$ are
determined as
\begin{equation}
 {\rm Re}~u_{\rm D(EX)}=F_{\rm V}(E,\rho)v_{\rm D(EX)}(s),\
 {\rm Im}~u_{\rm D(EX)}=F_{\rm W}(E,\rho)v_{\rm D(EX)}(s).
 \label{e4}
\end{equation}
The radial $v_{\rm D(EX)}(s)$ interactions were kept unchanged, as derived from
the M3Y-Paris interaction \cite{An83}, in terms of three Yukawas. The
parameters in Eq.~(\ref{e3}) were adjusted iteratively until $U_0(E,\rho)$,
given by Eq.~(\ref{e2}), agrees closely with the tabulated JLM results at each
energy \cite{Je77} (see Fig.~\ref{f1}).
The new density dependent interaction, dubbed hereafter as CDJLM interaction,
is then used in the DFM to calculate the complex OP for the \cc and
$^{6}$Li,$^{6}$He + $^{12}$C systems. From the results obtained for \cc system
(Fig.~\ref{f2}) we found that the mean-field absorption is surprisingly strong
and accounts up to 70\% of the total absorption. Since $W_{\rm F}$ does not
contain any contribution from numerous nonelastic reaction channels, the
imaginary \cc folded potential needs to renormalized by a factor $N_{\rm I}>1$
at all considered energies, while the renormalization factor of the real folded
potential remains close to unity (see right panel of Fig.~\ref{f2}). The
overall description of the elastic data and measured total reaction cross
section is very satisfactory which shows the reliability of the CDJLM
interaction.
Given an accurate prediction of $U_{\rm F}$, we have further estimated the
strength of polarization potential $\Delta V$ caused by the breakup effect in
the elastic $^{6}$Li,$^{6}$He + $^{12}$C, using the spline method from
Ref.~\cite{Kh95}. The effective (local) $\Delta V$ obtained for $^{6}$He turns
out to be much weaker than that obtained for $^{6}$Li, especially at high
energy (see Fig.~\ref{f3}). Such an unusual difference in the (breakup)
polarization potential $\Delta V$ for these two nuclei is being further
investigated.

\end{document}